\documentclass[11pt,a4]{article}
\usepackage{times,fancyhdr}
\usepackage[dvips]{graphicx}
\usepackage{amsmath}
\usepackage{amssymb}
\usepackage{array}
\usepackage{breakcites}
\usepackage[normalsize,bf]{caption}
\usepackage{color}
\usepackage[yyyymmdd,hhmmss]{datetime}
\usepackage{fancyhdr}
\usepackage{fancyvrb}
\usepackage{hyperref}
\usepackage{longtable}
\usepackage{setspace}
\usepackage{framed}
\usepackage{titlesec}
\usepackage{titling}
\usepackage{xcolor}
\usepackage{authblk}
\usepackage{booktabs}
\usepackage{ragged2e}
\usepackage{lineno}
\usepackage{comment}

\fancyheadoffset[R]{0pt}
\fancyfootoffset[R]{0pt}
\definecolor{orange}{rgb}{1.0,0.5,0.0}
\definecolor{aqgr}  {rgb}{0.0,1.0,0.6} 
\definecolor{dred}  {rgb}{0.8,0.0,0.0}
\definecolor{dchgd} {rgb}{0.0,0.5,0.0}
\definecolor{figdr} {rgb}{1.0,1.0,1.0} 
\definecolor{coldr} {rgb}{1.0,0.8,0.0} 
\definecolor{colop} {rgb}{0.3,0.7,1.0} 
\definecolor{colok} {rgb}{0.0,0.5,0.0} 
\definecolor{colcg} {rgb}{0.7,0.0,0.0}

\newcolumntype{L}[1]{>{\raggedright\arraybackslash\hspace{0pt}}p{#1}}
\newcolumntype{P}[1]{>{\centering\arraybackslash}p{#1}}
\newcommand\bigttl[1]{\title{\bfseries{#1}}}

\newcommand\pardr[1]{\colorbox{coldr}{
  \textcolor{black}{\textit{#1}}}}
\newcommand\parop[1]{\colorbox{colop}{
  \textcolor{white}{\textbf{#1}}}}
\newcommand\parok[1]{\colorbox{colok}{
  \textcolor{white}{\textsc{#1}}}}
\newcommand{\Marker}[2]{\ifnum\coloron=1
   \colorbox{#1}{#2}\else#2\fi}

\newcommand\neudef[1]{\textcolor{black}{\textbf{\textit{#1}}}}
\def\afhead{0}
\def\coloron{0}

\bigttl{ 
  A conceptual model for the evo-devo role of transposable
  elements and its implications for the ageing phenomenon
}

\author{
{\normalsize \textbf{Alessandro Fontana}}\\
{\normalsize \texttt{fontalex00@gmail.com}}}
\vspace{-1.0cm} \date{}

\makeatletter
\renewcommand\@biblabel[1]{}
\makeatother

\begin{document}

\pretitle{%
\begin{center}\LARGE
\vskip -2.2cm
\rule{\textwidth}{2.0pt}
\par
\vskip 0.5cm
}
\posttitle{
\par
\rule{\textwidth}{2.0pt}
\end{center}
\vskip -0.0cm
}
\maketitle

\vspace*{-2.5cm}
\begin{center}
\begin{tabular}{P{5.0cm} P{0.0cm} P{0.0cm}}
\end{tabular}
\end{center}
   
\vspace*{0.0cm}
\begin{abstract}
\normalsize
\if\afhead1 {\parop{xxxx}} \fi
The Evolvable Soma Theory of Ageing is a recently proposed model that frames development as a continuous process of change accompanying organisms throughout the lifespan. This process is driven by developmental genes which encode epigenetic changes on target cells, whereas ageing reflects the expression of late-acting modifications, that are subject to ongoing evolutionary optimisation and function as somatic ``experiments'' to explore phenotypic novelty. In this work we examine the role of transposable elements in the model. Our proposal acknowledges that these elements facilitate the expansion and diversification of gene regulatory networks by providing transcription factor binding sites. To minimise disruption, their regulatory activity is tightly repressed by epigenetic mechanisms during early development, which may be progressively released by genetically driven, age-associated epigenetic changes in later life, thereby contributing to transcriptional pseudorandomness and ageing-associated phenotypes. Within this framework, transposable elements are integrated into a unified view of evolution, development and ageing, providing a conceptual basis for their dual role in regulatory innovation and age-related decline.
\end{abstract}

\tableofcontents

\clearpage

\section{Introduction}

\if\afhead1 {\parop{ageing, mechanistic causes}} \fi
Numerous mechanistic theories have been proposed to explain ageing. Prevailing stochastic models attribute it to the progressive accumulation of damage arising from environmental insults and endogenous processes, in a manner analogous to the wear and tear of mechanical systems. Consistent with this view, a range of ageing hallmarks has been identified, including genomic instability, telomere attrition, epigenetic alterations, and loss of proteostasis \cite{LopezOtin13}. However, it remains debated which of these processes act as primary drivers of ageing and which represent secondary consequences or downstream effects.
 
\if\afhead1 {\parop{ageing and development}} \fi
If the biological basis of ageing remains unresolved, its relationship with development and evolution is even more perplexing. Prevailing frameworks implicitly divide the lifespan into two phases: a developmental phase, driven by a finely tuned genetic programme, followed by a maintenance phase characterised by progressive deterioration, giving rise to the ageing phenotype. This introduces a sharp conceptual boundary between two phases assumed to be mechanistically distinct. The distinction echoes engineering analogies, in which construction is followed by maintenance. From an epistemological perspective, however, such a dichotomy is unsatisfactory, as it imposes an abrupt and somewhat arbitrary transition within the lifespan.

\if\afhead1 {\parop{ageing and evolution}} \fi
As far as evolution is concerned, the prevailing view is that ageing is non-adaptive, persisting because the force of natural selection declines with age, leading to reduced investment in cellular maintenance in favour of early-life reproductive success, or arising from trade-offs involving genes with pleiotropic effects \cite{Kirkwood00}. In contrast, alternative hypotheses propose that ageing may itself be adaptive and selected for its benefits to either individuals, to groups, or to evolutionary dynamics. A. Weismann suggested that death evolved to promote generational turnover, consistent with group selection arguments \cite{Weismann82age}, while others have argued that limited lifespan prevents older individuals from dominating the gene pool, thereby enhancing evolutionary adaptability \cite{Goldsmith08}.

\if\afhead1 {\parop{transposable elements}} \fi 
In this work, we address the problem using a model that was proved useful in interpreting a range of biological phenomena, from which emerges a recently proposed hypothesis concerning the nature of ageing. Within this framework, we investigate the role of transposable elements, a class of genetic elements widely distributed across the genomes of many species and hypothesised to contribute to evolution, ageing, and disease. Despite their abundance in the genomes of many species and their potential importance in shaping genome regulation and dynamics, their broader biological significance remains only partially understood. This work aims to propose a conceptual framework for the evo-devo role of transposable elements, for which empirical validation of specific predictions will be required.

\if\afhead1 {\pardr{structure}} \fi 
This paper is organised as follows. After this introduction, Section 2 outlines the current understanding of the evo-devo implications of transposable elements, summarising the main experimental and comparative evidence. Section 3 introduces the developmental framework and the associated hypothesis on the nature of ageing. Section 4 outlines a model for the role of transposable elements in development and ageing within this framework. Sections 5 and 6 then present two alternative models for the evolutionary dynamics of transposable elements. Section 7 discusses the broader implications of these results in the context of development and evolution, and Section 8 concludes by summarising the main findings and outlining directions for future research.

\section{Evidence for the evo-devo role of transposable elements}
\label{sec:trelevod}

\if\afhead1 {\parok{TE, genome struct}} \fi
Transposable elements (TE) \cite{McClintock50} are ubiquitous components of eukaryotic genomes \cite{Chang22}, comprising substantial fractions of genomic DNA (for example, approximately 40\% in humans and up to 90\% in maize). Across many species the number and overall structure of protein-coding genes are broadly conserved. The principal genomic differences instead lie in the mean distance between adjacent genes, which scales with genome size and largely reflects extensive insertions of transposable elements within intergenic regulatory regions. A clear example is provided by mammals such as bats, dogs, and marsupials, which differ substantially in genome size (from approximately 2.0–2.3 Gb in bats, to 2.4 Gb in dogs, and 3.5–3.7 Gb in marsupials \cite{Gregory05}) while retaining comparable repertoires of protein-coding genes. 

\if\afhead1 {\parop{TE, degrees of autonomy}} \fi
It is worth noting that in many species, including humans, only a very small subset of transposable elements retains the full sequence complement required for autonomous transposition. In humans, active autonomous elements constitute less than 0.05\% of the genome and are represented primarily by a small number of retrotransposition-competent LINE-1 elements \cite{Beck15}. A larger fraction of the genome, approximately 10--11\%, consists of non-autonomous elements such as Alu and SVA sequences, which remain potentially mobile by exploiting the enzymatic machinery encoded by active LINE-1 elements \cite{Mills07}. By contrast, the majority of TE-derived sequences, representing roughly 33--35\% of the human genome, are truncated or inactivated by mutations and deletions and are no longer capable of transposition. This genomic architecture likely reflects the long-term co-evolutionary dynamics with their hosts, in which transposition potential becomes progressively restricted while residual TE-derived sequences accumulate throughout the genome \cite{Kazazian04}.

\if\afhead1 {\parok{TE, germline activity}} \fi
Transposable elements can be active in both germline cells and during early embryogenesis, stages characterised by extensive epigenetic reprogramming and a transient relaxation of transposon silencing mechanisms \cite{Goodier16}. During these windows of developmental plasticity, global DNA demethylation and chromatin remodelling can reduce host-mediated repression, allowing increased TE transcription and, in some cases, mobilisation. Such activity is tightly regulated but not completely suppressed, and can contribute to insertional mutagenesis as well as the generation of somatic and germline mosaicism. These dynamics are particularly relevant in mammals, where early embryonic development and primordial germ cell formation are associated with heightened sensitivity to TE activation.

\if\afhead1 {\parop{TE, somatic}} \fi
TE activity is not restricted to the germline or early development but can also be observed in somatic tissues, where they contribute to genomic mosaicism. A prominent example is provided by studies of LINE-1 retrotransposition in the mammalian brain, where somatic insertions have been reported in neuronal precursor cells and differentiated neurons. As demonstrated in \cite{Muotri05}, LINE-1 elements can mobilise during neurogenesis, leading to genetically distinct neuronal populations within the same individual. This somatic TE activity is documented in a wide range of organs and tissues \cite{Wang25somat}, suggesting that retrotransposition is not confined to the germline but can occur in specific developmental or cellular contexts.

\if\afhead1 {\parop{TE, roles}} \fi
Originally described as ``control elements'' \cite{McClintock50}, reflecting their potential role in regulating gene expression, transposable elements were subsequently reframed as ``selfish DNA'' or ``genomic parasites'' \cite{OrgelCrick80}, emphasising their capacity to propagate within genomes largely independently of host fitness. However, this view has been substantially revised in light of accumulating evidence that these elements are frequently co-opted by host genomes. In particular, they can provide regulatory sequences such as promoters, enhancers, and transcription factor binding sites (TFBS) \cite{Rebollo12}, thereby contributing to the evolution of gene regulatory networks \cite{Oliver10}.

\section{ESTA hypothesis}
\label{sec:model}

\begin{figure*}[t] \begin{center} \hspace*{-1.00cm}
\includegraphics[width=18.60cm]{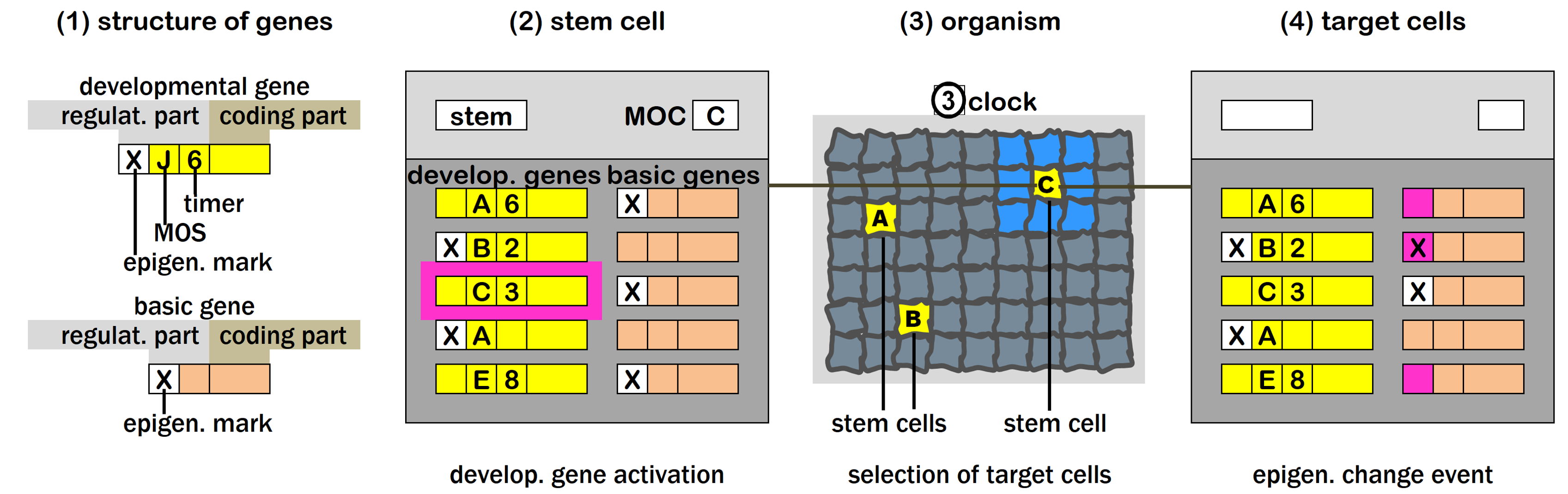}
\caption{
\if\afhead1 {\parop{figurex}} \fi
\if\afhead1 {\parok{caption}} \fi
The genome is divided into basic and developmental genes (panel 1). Developmental genes encode change events that reshape the epigenetic landscape of specific target cells. In this example, a developmental gene with MOS=C and timer=3 is activated in stem cell with MOC=C when the global clock reaches value 3 (panels 2 and 3). Upon activation, the gene specifies both the target cells and the epigenetic modifications to be imposed on their initial state, leading to the structural disabling or enabling of certain (basic or developmental) genes (panel 4). Depending on the encoded instructions, the resulting downstream outcome may be a change in cellular behaviour, a proliferation event, or an apoptosis event.
}
\label{events}
\end{center} \end{figure*}

\if\afhead1 {\parok{model description}} \fi
The ageing theory examined here is grounded in a model of development originally introduced in \cite{Font08}, which has been shown able to generate complex structures in computer simulations, and interpret a range of biological phenomena \cite{Font23a}. Fig.~\ref{events} provides an illustration of the key model elements. Phenotypes are composed of \neudef{normal cells} and (fewer) \neudef{stem cells} operating in synchrony with a \neudef{global clock}, an abstract representation of developmental time. Each stem cell is characterised by a \neudef{Master Organisation Code (MOC)} that captures its positional and organisational identity. The genome encodes two distinct classes of genes: \neudef{basic genes}, which produce structural components, and \neudef{developmental genes}, which orchestrate development. 

\if\afhead1 {\parop{model description}} \fi
Genes of both categories consist of a regulatory part and a coding part. The regulatory part contains a field that can bear an \neudef{epigenetic mark}: once the mark is present, the gene is structurally disabled and excluded from the network. The regulatory part of developmental genes contains also a \neudef{Master Organisation Sequence (MOS)}, which determines compatibility with a specific MOC, and a \neudef{timer}, which specifies the temporal window of gene activation by matching the state of the clock. When both spatial (MOS–MOC) and temporal (timer–clock) conditions are satisfied, a \neudef{change event} is triggered, which adds or removes the epigenetic marks of genes in target cells, thereby inducing differentiation, proliferation, or apoptosis. Development is thus conceived as a sequence of spatio-temporally coordinated change events that collectively drive tissue specification and organ formation.

\if\afhead1 {\parok{epigen marks on dev genes}} \fi
This version of the model introduces a novel feature compared with previous formulations, namely the inclusion of epigenetic marks on developmental genes in addition to those on basic genes. These marks inhibit gene activation and can be added or removed through the action of specific developmental genes, in a manner analogous to the regulation of basic genes. This mechanism introduces a third condition for activation of developmental genes, alongside MOS and timer control. In Section~\ref{sec:treltfbs}, we will examine how this additional regulatory mechanism in developmental genes can model the influence of transposable elements on the ageing process.

\if\afhead1 {\parok{biological interpretation}} \fi
The distinction between basic and developmental genes is primarily functional: the former control cellular behaviour within a given epigenetic state, the latter direct state changes. While basic genes are hypothesised to be implemented through protein-coding genes, developmental genes may encompass a broader range of genetic elements, including protein-coding genes as well as long non-coding RNAs, which are increasingly recognised as key regulators of epigenetic states \cite{Rinn14, Perry16}. Epigenetic regulation comprises a diverse set of chemical modifications that coordinate gene activity across time and space, including the writing, erasing, and interpretation of the ``histone code'' \cite{Jenuwein01}, as well as the establishment and removal of DNA methylation marks. These mechanisms, whose genetic and molecular pathways are now beginning to be elucidated \cite{Xu25epigen}, lie at the core of the developmental process.

\begin{figure*}[t] \begin{center} \hspace*{-0.80cm}
\includegraphics[width=18.20cm]{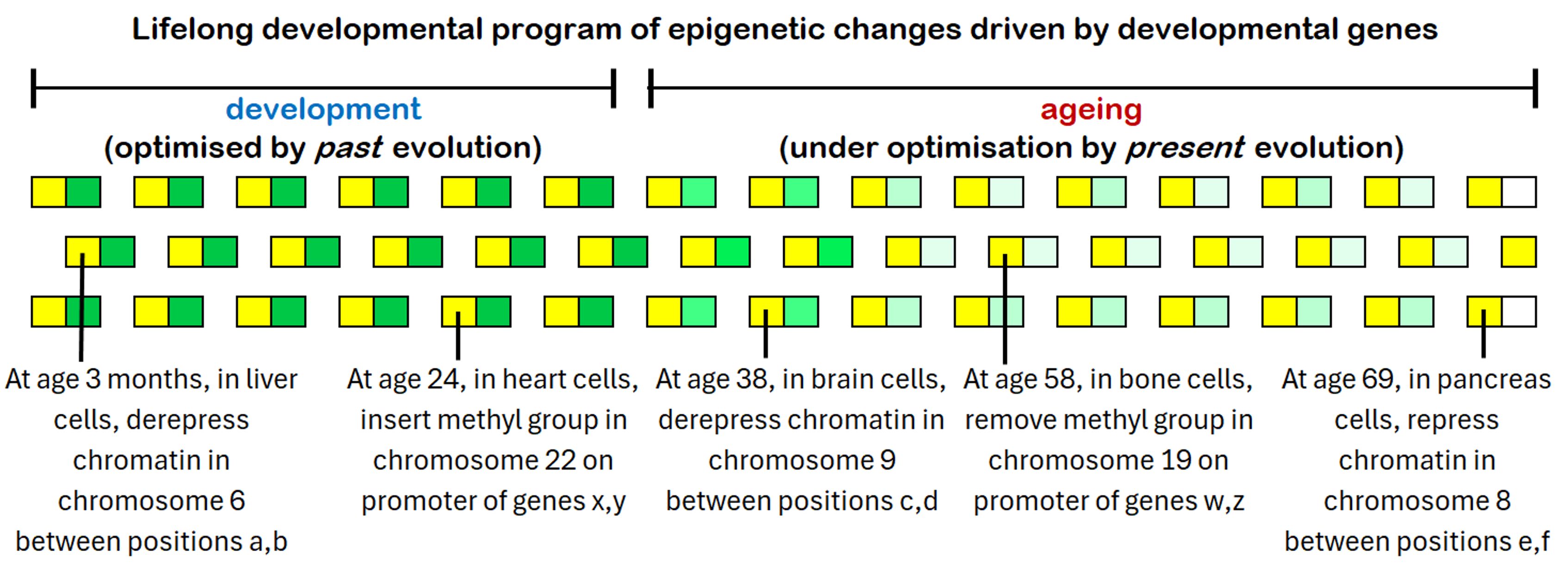}
\caption{
\if\afhead1 {\parop{figurex}} \fi
\if\afhead1 {\parop{caption}} \fi
Lifelong developmental programme. Development and ageing form a continuous, genetically regulated process, driven by developmental genes whose activation is controlled by their associated timers. Developmental genes trigger epigenetic change events (some examples are reported). Genes that are activated after reproduction are progressively less optimised (optimisation degree represented by green shade), functioning as evolution ``experiments''. Development represents the early part of the developmental programme, optimised by past evolution, whereas ageing is the late part of the programme, under optimisation by present evolution.}
\label{devprog}
\end{center} \end{figure*}

\if\afhead1 {\parop{development and pseudo-randomness}} \fi
In our model development and ageing form a continuous, genetically regulated process: the developmental programme proceeds through a sequence of \neudef{developmental steps}, each associated with a specific clock value and a set of coordinated change events (Fig.~\ref{devprog}). After reproduction, the impact of change events on the evolutionary fitness decreases as the time of their occurrence (specified by the associated timer value) increases. In this context, ageing can be explained by noting that developmental genes activated after the age of reproduction contribute less to fitness and are therefore subject to weaker evolutionary optimisation. Although their effects are genetically encoded and deterministic, they are less optimised by evolution, raising the probability of detrimental outcomes.

\if\afhead1 {\parop{ESTA intro}} \fi
Accordingly, the \neudef{Evolvable Soma Theory of Ageing (ESTA)} \cite{Font24a} posits that development is governed by a portion of the genetic programme already optimised by past evolution, whereas ageing emerges from the action of late-acting genes that are still under evolutionary optimisation. These post-reproductive genetic modifications function as somatic ``experiments'' that often impair physiological function but can, on occasion, generate beneficial variation on which natural selection may act across generations. From this perspective, ageing is not merely a process of decline, but can be understood as \textit{evolution in action}. 


\if\afhead1 {\clearpage} \fi

\section{Model of transposable elements in development and ageing}
\label{sec:treltfbs}

\begin{figure*}[t] \begin{center} \hspace*{-0.20cm}
\includegraphics[width=17.00cm]{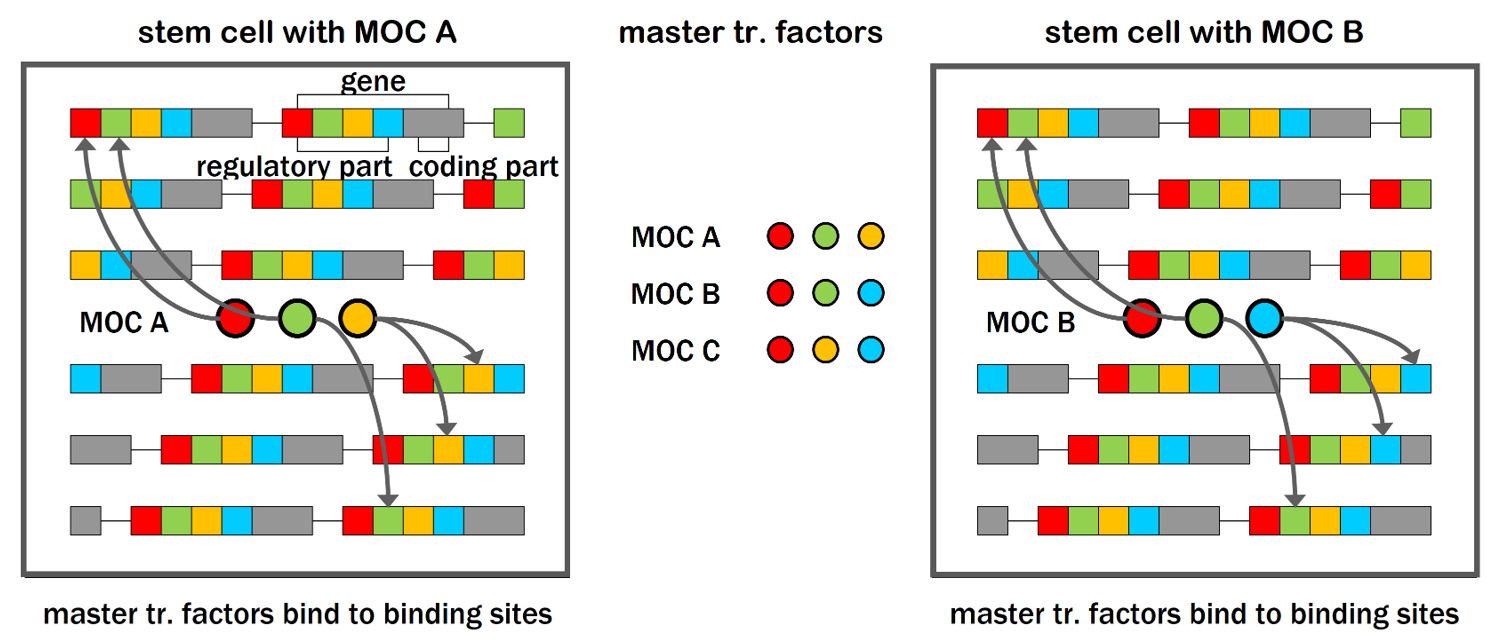}
\caption{
\if\afhead1 {\parop{figurex}} \fi
\if\afhead1 {\parok{caption}} \fi
Biological implementation of MOC and MOS. The MOC (labelled by letters) is hypothesised to represent a specific combination of master transcription factors, with each MOC corresponding to a distinct set of factors. These transcription factors bind to matching regulatory sequences in the genome, which collectively constitute the MOS. Their interaction initiates regulatory cascades that ultimately shape epigenetic change events in target cells.
}
\label{factors}
\end{center} \end{figure*}

\if\afhead1 {\parok{factors, introduction}} \fi
Our developmental framework assigns a central role to the interaction between MOC and MOS, which is required to trigger change events. In this section, we make this proposal more concrete by outlining a low-level biological implementation of these two key elements. MOC are hypothesised to be composed of specific combinations of \neudef{master transcription factors} (Fig.~\ref{factors}), while MOS correspond to the TFBS to which these factors bind. Master transcription factors occupy the highest levels of the cellular regulatory hierarchy: by binding to their corresponding TFBS, they initiate cascades of gene activations that ultimately shape epigenetic change events in target cells.

\if\afhead1 {\parok{core hypothesis: TE $<->$ MOS}} \fi
Our hypothesis is that transposable elements may constitute, or contribute to, the TFBS of developmental genes, corresponding in our framework to MOS elements. This possibility is consistent with growing evidence that a substantial fraction of TFBS in eukaryotic genomes derives from TE sequences \cite{Sundaram14}, although the extent and functional significance of this contribution remain under investigation. In particular, transposable elements have been implicated in the regulation of developmental genes \cite{Lowe07}, highlighting their potential role in shaping developmental regulatory networks. Within the present framework, TE-derived regulatory sequences are therefore hypothesised to participate in the cell-specific control of developmental genes and to the orchestration of change events throughout the lifespan.

\if\afhead1 {\parop{TEs influence epigenetics}} \fi
Transposable elements are known to influence DNA methylation patterns and chromatin organisation, thereby contributing to the regulation of gene expression and genome accessibility \cite{Lippman04}. Through their interactions with epigenetic silencing machinery, they can promote the formation of heterochromatin, act as boundaries between chromatin domains, and affect the activity of nearby genes. More recently, TE-derived sequences have also been implicated in the organisation of higher-order chromatin architecture, including enhancer-promoter interactions and three-dimensional genome folding \cite{Lawson23}. Through these effects, transposable elements  can participate in the establishment and modulation of large-scale epigenetic states, in agreement with our hypothesis. 

\begin{figure*}[t] \begin{center} \hspace*{-1.00cm}
\includegraphics[width=18.60cm]{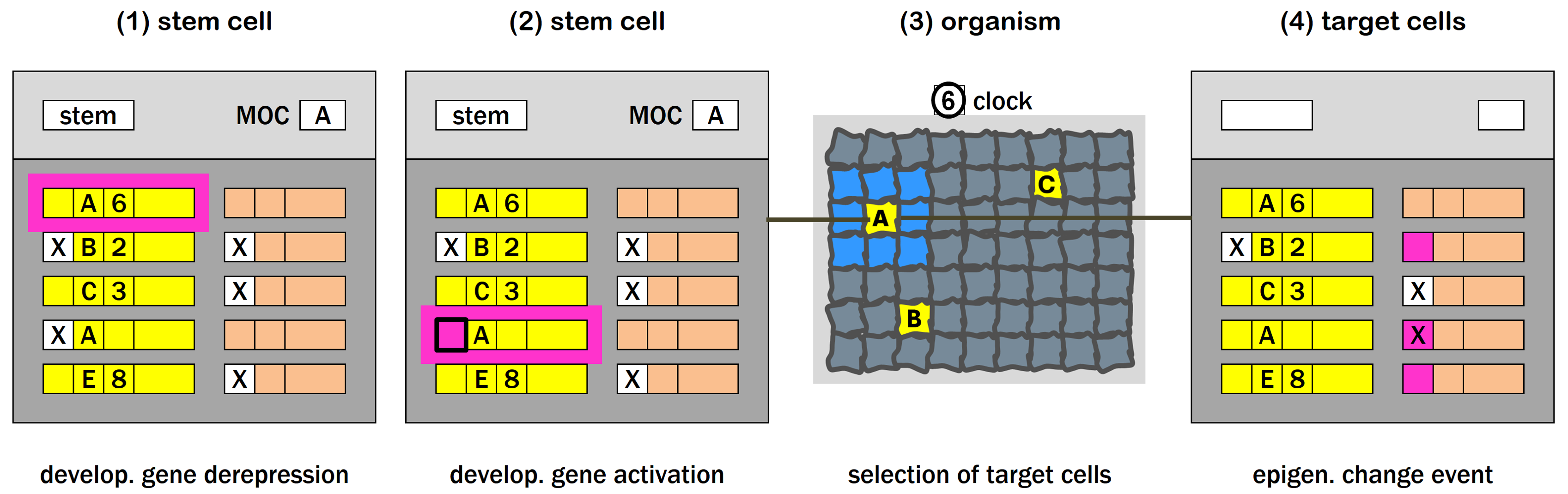}
\caption{
\if\afhead1 {\parop{figurex}} \fi
\if\afhead1 {\parop{caption}} \fi
Activation of TE-regulated developmental genes during ageing. First, the action of a developmental gene (panel 1) leads to the epigenetic derepression of a second developmental gene (through the removal of its associated methylation mark), whose timer-independent activation (panel 2) leads to an epigenetic change event (panel 4).}
\label{ageact}
\end{center} \end{figure*}

\if\afhead1 {\parok{ESTA ageing recap}} \fi
Building on the hypothesised role of transposable elements, we can now refine the ESTA model to explicitly incorporate their contribution to the developmental programme. The original formulation of ESTA posits that developmental genes can be activated throughout the entire lifespan, with the timing of their activation in specific target cells determined by their associated timer values. Developmental genes scheduled for activation during the post-reproductive period are expected to be less optimised by evolution, and their activation contributes to failed somatic experiments and to the phenotypic manifestations of ageing, including physiological decline and age-associated diseases. In this work we introduce an additional layer of regulation of developmental genes through their associated methylation marks, thereby providing a mechanistic framework through which transposable elements may contribute to the ageing process.

\if\afhead1 {\parok{gradual release of MOS methylation}} \fi
We propose that the MOS fields associated with non-optimised developmental genes are initially methylated at the onset of development. This step is needed because these genes are not fully refined and may therefore exert potentially disruptive or deleterious effects if expressed prematurely. After reproduction, this methylation state is progressively relaxed through the action of other developmental genes, gradually releasing control and allowing these latent genes to become active, in a period of reduce evolutionary pressure, when consequences of failed experiments is less impactful. This mechanism introduces an additional regulatory layer governing evolutionary experimentation, which constitutes the core feature of the ESTA model.

\if\afhead1 {\parop{some evidence}} \fi
The proposed model is broadly consistent with experimental evidence. A recent study \cite{Wasserzug22} showed that ageing oocytes progressively lose heterochromatin-associated epigenetic marks, leading to retrotransposon activation, increased DNA damage, and impaired oocyte maturation. Notably, disrupting heterochromatin in young oocytes reproduced these defects, whereas restoring heterochromatin or inhibiting retrotransposon activity partially rescued oocyte function. Another study \cite{Zeng26} reported widespread age-associated changes in chromatin organisation and transposable element regulation in the ageing mouse brain, together with altered accessibility of regulatory regions. Although neither study directly tests the present hypothesis, both are consistent with a model in which progressive relaxation of epigenetic control over transposable elements contributes to ageing-related chromatin remodelling, gene dysregulation and functional decline.

\if\afhead1 {\parok{de-methylation in standard model}} \fi
In the prevailing view, transposable elements are methylated during early development because their potential to compromise genome integrity has driven the evolution of multiple host defence mechanisms that suppress their activity. As reviewed in \cite{Chuong17}, these defence systems operate at several levels, including epigenetic silencing via DNA methylation and histone modifications, as well as small RNA–mediated pathways such as the piRNA system, which specifically targets TE transcripts for repression. Together, these mechanisms are hypothesised to suppress TE activity and limit their potentially deleterious effects on genome stability and gene regulation.

\if\afhead1 {\parop{methylation to prevent binding not jump}} \fi
However, as noted in Section~\ref{sec:trelevod}, the vast majority of transposable elements in most genomes are no longer capable of either autonomous or non-autonomous mobilisation. In this light, it can be argued that the relevant activity to be controlled is not transposition per se, but rather the regulatory potential of TE-derived sequences. Their capacity to function as cis-regulatory elements is therefore likely to be of greater biological relevance, since unregulated transcription factor binding could trigger inappropriate activation of developmental genes. This interpretation is consistent with the model proposed here, in which epigenetic silencing primarily serves to control the TE-derived regulatory activity of the developmental programme.

\section{TE evolutionary dynamics: standard model}

\if\afhead1 {\parok{chicken-and-egg problem}} \fi
In the previous section, we proposed a model for the developmental (``devo'') role of transposable elements. Here, we turn to the evolutionary (``evo'') dimension, and try to address the question of how the regulatory function of transposable elements may have emerged during evolution. More specifically, the issue concerns the origin of the highly precise correspondence between transcription factors and their binding sites. This problem reflects a broad ``chicken-and-egg'' dilemma in regulatory evolution: transcription factors bind DNA with high specificity, yet it remains unclear how matching factors and binding sites initially emerge and become integrated into functional regulatory networks. 

\if\afhead1 {\parok{TE are TFBS-enriched}} \fi
A widely supported model, which for convenience we refer to as the \neudef{standard model}, \cite{Feschotte08}, proposes that successive colonisation waves have dispersed throughout the genome TE sequences, which are naturally enriched in transcription factor binding motifs or proto-regulatory sequences. Many TE families carry pre-existing cis-regulatory features (such as promoter elements, enhancers, and short TFBS motifs) which can be readily adapted to host regulatory contexts. For example, endogenous retroviruses often harbour binding motifs for pluripotency factors (e.g. OCT4, SOX2, NANOG), while other elements contain sites for factors such as p53 or CTCF \cite{Kunarso10}. Following insertion, these motifs can be refined by subsequent mutations and selection, giving rise to functional binding sites with context-specific activity. 

\if\afhead1 {\parok{colonisation waves}} \fi
Successive waves of transposable element expansion in genomes have long been associated with periods of increased evolutionary innovation, potentially coinciding with the origin of new lineages and contributing to the emergence of novel body plans and species-specific traits \cite{Oliver10}. Although direct causal links are not yet established, comparative genomic studies have shown that bursts of TE activity often coincide with episodes of regulatory innovation and extensive rewiring of gene regulatory networks, particularly those involved in development \cite{Chuong17}. Within this perspective, the dispersal of new TE families throughout the genome may provide the raw regulatory material upon which natural selection acts, facilitating the diversification of developmental programmes during the evolution of the tree of life.

\if\afhead1 {\parok{non-optimised regul. and coding parts}} \fi
This standard model is broadly compatible with the hypothesis that transposable elements are the key component of MOS elements in our framework. It is worth noting that, while such TE-enriched TFBS are still undergoing optimisation during evolution, the coding sequences they control (responsible for mediating epigenetic changes in target cells) are also likely to be partially optimised. In such a scenario, both the regulatory input (binding specificity and affinity) and the downstream effector functions may be only partially tuned, leading to imprecise or context-inappropriate gene activation, with potentially deleterious consequences for cellular function. This provides a rationale for the epigenetic silencing of such developmental genes during early development, while their gradual release after reproduction allows their expression during a period of reduced evolutionary pressure, where unsuccessful ``experiments'' incur a lower fitness cost.


\section{TE evolutionary dynamics: alternative model}

\begin{figure*}[t] \begin{center} \hspace*{-0.20cm}
\includegraphics[width=17.00cm]{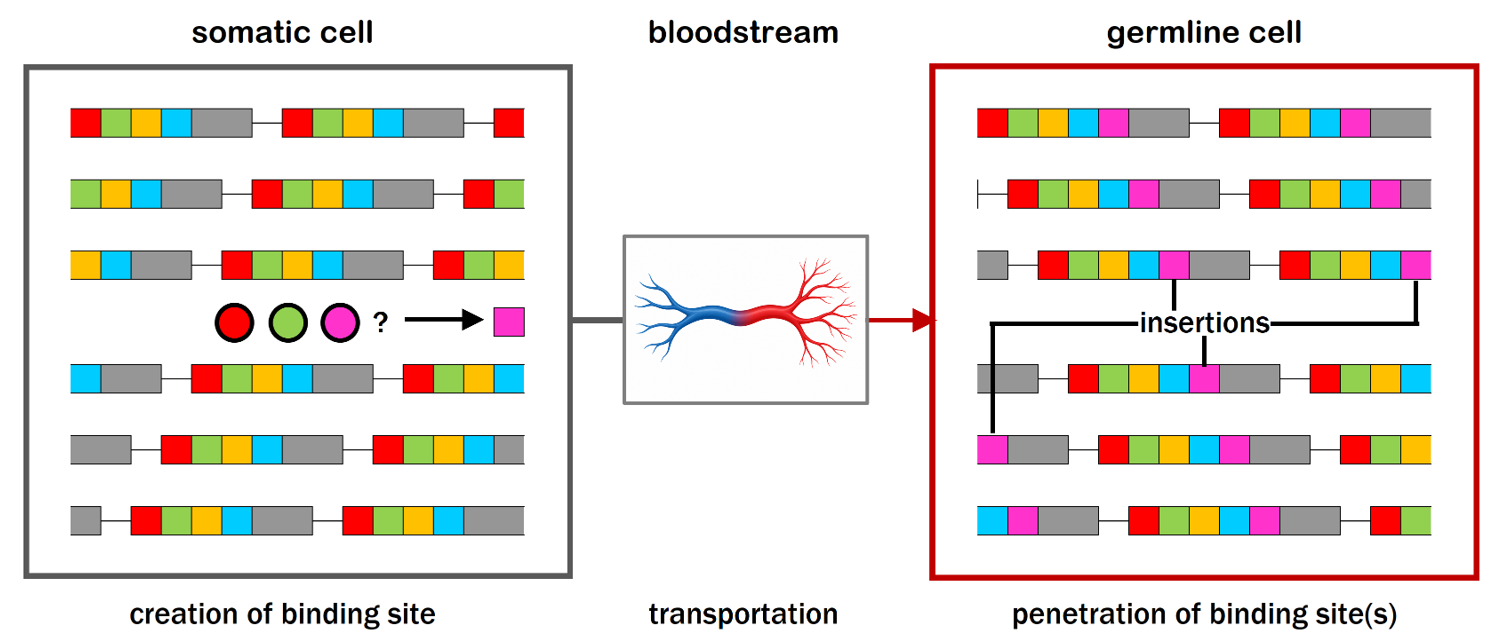}
\caption{
\if\afhead1 {\parop{figurex}} \fi
\if\afhead1 {\parop{caption}} \fi
Germline penetration of binding sites. In the left panel, the genome is unable to trigger a change event because it lacks binding sites compatible with the purple transcription factor. A novel binding site for this factor is generated, leaves the cell and through the bloodstream reaches the germline, where it is incorporated into the regulatory part of developmental genes alongside pre-existing binding sites. The modified germline genome gives rise to the somatic genome of subsequent generations. After evolutionary optimisation of the associated gene, the purple transcription factor can trigger a change event and contribute to development.
}
\label{gpenet}
\end{center} \end{figure*}

\if\afhead1 {\parop{GP, formulation}} \fi
In this section, we present an \neudef{alternative model} for TE evolutionary dynamics, which includes less-canonical elements. In this scenario, transcription factors arise first, and their corresponding binding sites are generated subsequently. Consider a situation in which a novel transcription factor is expressed for the first time in the evolutionary history of a species, during the development of a given individual in a specific somatic cell (Fig.~\ref{gpenet}, left panel). In the absence of compatible binding sites in the genome of that cell (inherited from the parental germline), the transcription factor is unable to exert regulatory effects and its functional potential remains unrealised.

\if\afhead1 {\parop{GP, formulation}} \fi
One possible way to recover this regulatory potential is to envisage the transcription factor acting as a kind of mould for the generation of a compatible binding site within the somatic cell, analogous to a key shaping its own lock. Following its formation, the binding site would be transferred to the germline, for example via circulation, to be incorporated, potentially in multiple copies, into the germline genome subsequently inherited by future generations (Fig.~\ref{gpenet}, right panel). Once sufficient evolutionary optimisation has occurred, the interaction of the transcription factor with its compatible binding site can take place reliably, enabling regulated gene activation and thereby fostering the emergence of novel developmental functions.

\if\afhead1 {\parop{GP in computer simulations}} \fi
This alternative model of TE evolutionary dynamics was originally inspired by results obtained in computer simulations. In these simulations, a related mechanism was introduced to increase the efficiency of the evolutionary process. The procedure, termed \neudef{germline penetration} \cite{Font10c, Font12b}, assumes that MOS sequences compatible with MOC elements generated during an individual’s development can be copied into developmental genes in the genome of subsequent generations. After integration, these genes are initially maintained in an inactive state through an epigenetic mark, which prevents their premature activation while their coding regions may still be insufficiently optimised.

\if\afhead1 {\parop{somatic activity and transfer}} \fi
The occurrence of somatic TE activity, particularly in the brain \cite{Muotri05}, is compatible with this alternative model. Within the present framework, such activity can be interpreted not merely as stochastic genomic instability, but as part of a broader exploratory mechanism capable of generating local regulatory variation during development. A subsequent step would involve the transfer of such sequences from somatic cells to the germline. One hypothetical route for this process involves circulating nucleic acids or endogenous viral elements released from cells in association with proteins, lipids, or extracellular vesicles \cite{Thierry16}. These circulating materials may include retroelement-derived sequences, including LINE-1-related elements \cite{Koh14LINE1}, potentially providing a mechanism for the dissemination of new regulatory sequences.

\if\afhead1 {\parop{SMGT process}} \fi
Transposable elements have also been implicated in a process called sperm-mediated gene transfer (SMGT), whereby spermatozoa can take up exogenous DNA and deliver it to oocytes at fertilisation \cite{Smith05}. Following reverse transcription, such sequences could, in principle, integrate into the germline genome and become heritable. Consistent with this possibility, genetic material released from somatic cells has been reported to be detectable in the offspring of exposed individuals \cite{Cossetti14}. The SMGT process  appears conceptually consistent with the alternative model, although its frequency, underlying mechanisms, and biological significance remain uncertain.

\if\afhead1 {\parop{male bias}} \fi
An additional observation concerns sex-specific differences in TE activity. Transposable elements appear to be more active in the male germline. Elevated expression and mobilisation of elements such as LINE-1 have been reported in male germ cells and during spermatogenesis \cite{Bao12male, Rebollo12}. This pattern may reflect fundamental differences in germline biology between the sexes: male germ cells undergo continuous mitotic divisions throughout life, thereby increasing the opportunities for TE mobilisation and insertion events, whereas the female germline becomes largely mitotically quiescent after embryogenesis, substantially limiting such opportunities. These observations are broadly consistent with the alternative model, which predicts a greater contribution of the male germline to the generation and propagation of evolutionary innovations.

\if\afhead1 {\clearpage} \fi

\section{Discussion}

\if\afhead1 {\parok{evo-devo role}} \fi
The role of transposable elements in genome evolution is well established. By reshaping regulatory networks, altering gene expression, and generating novel genetic variation, they provide an important source of evolutionary innovation upon which natural selection can act. The contribution of the present work is to extend this perspective by explicitly integrating development into the evolutionary equation. Rather than viewing these elements solely as generators of evolutionary novelty at the population level, the proposed framework emphasises their potential role within a temporally organised developmental programme, where TE-mediated regulatory changes may influence the sequential modulation of life stages across the lifespan and the emergence of ageing-related phenotypes.

\if\afhead1 {\parop{developmental biology}} \fi
We argue that some limitations of current theories of ageing may reflect the absence of a unified, generative understanding of developmental processes. Developmental biology, particularly beyond early embryogenesis, has historically received less attention and fewer resources than fields such as molecular genetics, neuroscience, or oncology \cite{Rogers24}. As a result, although many molecular and cellular mechanisms of development have been extensively characterised, our ability to design models that reproduce the development of complex phenotypes remains limited. We suggest that explicitly incorporating the ``devo'' component into evolutionary models provides a more comprehensive conceptual framework, better suited to capturing biological complexity and, consequently, to explaining the origin of ageing.

\if\afhead1 {\parop{timer vs epigen mark}} \fi
The effect of transposable elements on the ageing phenotype is hypothesised to be mediated by the gradual, genetically controlled removal of epigenetic marks associated with developmental genes. These marks suppress gene activation until they are released at specific stages of the developmental programme. In our developmental framework, a similar effect could in principle be achieved through the timer field alone: assigning a timer value beyond the expected lifespan would effectively prevent gene activation. However, adjusting gene activity through epigenetic marks may provide a simpler and more flexible mechanism than fine-tuning timer values. In this sense, epigenetic regulation can be viewed as an efficient means of controlling the activation of developmental genes across the lifespan.

\begin{table}[ht!]
\centering 
\hspace*{-0.0cm} 
\begin{tabular}
{P{1.0cm}L{6.8cm}L{6.8cm}}
\hline  
& \textbf{Standard model} 
& \textbf{Alternative model} \\ \hline
(1) 
& colonisation of organism by external ``waves'' of transposable elements 
& creation in somatic cell of TE-derived TFBS compatible with given factor \\ 
(2) 
& propagation of ``raw'' elements in germline genome 
& transfer and dissemination of TFBS in germline genome \\ 
(3) 
& selection of TFBS and associated genes by evolution
& selection of TFBS and associated genes by evolution \\ 
\hline
\end{tabular}
\caption{
\textbf{Two models for TE evolutionary dynamics.}}
\label{twoscenarios}
\end{table}

\if\afhead1 {\parok{Weismann barrier, criticism}} \fi
The Weismann barrier \cite{Weismann92bar} describes the separation between germline and somatic cells, whereby genetic information is transmitted from germline to soma but not in the opposite direction. Although the Weismann barrier remains a central principle of modern evolutionary biology, several authors have argued that it may be less absolute than traditionally assumed. Work by Jablonka and collaborators suggests that environmentally induced epigenetic states can, under some conditions, persist across generations \cite{Jablonka05}. Shapiro has proposed that genomes should be viewed as actively modifiable ``read–write'' systems rather than passive repositories of information \cite{Shapiro11}, while Müller and proponents of the Extended Evolutionary Synthesis have stressed the evolutionary importance of developmental processes \cite{Mueller10}. These perspectives suggest that soma-germline interactions may be more complex than originally envisaged.

\if\afhead1 {\parok{GP, component 1}} \fi
The alternative model can be viewed as comprising three main components: (1) the contribution of transposable elements to the generation of transcription factor binding sites in somatic cells, (2) the possibility of soma-to-germline transfer, and (3) the ``mould hypothesis''. Among these, the first component rests on the strongest empirical foundation. As already noted, a substantial body of evidence indicates that transposable elements can contribute regulatory sequences, and that TE-derived elements have been repeatedly co-opted into host gene regulatory networks during evolution. In this respect, the assumption that TE-derived sequences may contribute to MOS-like regulatory elements can be regarded as plausible and broadly consistent with current knowledge.

\if\afhead1 {\parop{GP, component 2}} \fi
The second component remains considerably more uncertain, although several observations suggest that restricted transfer of genetic material between these compartments may be possible under specific conditions. For instance, even within the standard TE-evo framework, transposable elements and endogenous retroviruses of viral origin must at some stage have accessed the germline to become heritable. While this is typically assumed to occur via infection of germ cells or early embryos, viral entry often begins in somatic tissues, making indirect routes involving somatic cells at least plausible \cite{Coffin97}. Additional findings, including extracellular nucleic acids, vesicle-mediated transfer, and sperm-mediated gene transfer, can further support the hypothesis of soma-to-germline communication. However, the frequency, mechanisms, and evolutionary relevance of such processes remain unresolved.

\if\afhead1 {\parop{mould hypothesis}} \fi
Regarding the third component, while binding motifs are short (6–10 base pairs \cite{Stewart12}) and can arise relatively frequently, the emergence of multiple specific TFBS remains only partially understood. The mould hypothesis requires a cellular context in which transcription factors and transposable elements are simultaneously present, potentially allowing (through mechanisms that remain to be elucidated) the generation of binding sites compatible with the factor. As transcription factors are typically expressed in specific somatic contexts, whereas transposable elements are present in both somatic and germline cells, such interactions are expected to occur in somatic cells.  

\if\afhead1 {\parop{model comparison}} \fi
Table~\ref{twoscenarios} compares the two models of TE evolutionary dynamics proposed. In the standard model, novel regulatory sequences arise through external waves of TE colonisation, followed by their propagation in the germline genome and subsequent co-option and refinement by evolution. In the alternative model, the process begins with the generation of compatible binding sites in somatic cells, which are then transferred to and propagated within the germline before undergoing evolutionary optimisation. Thus, while both models share a final stage of evolutionary tweaking and selection, they differ in the proposed origin of novel TFBS: externally introduced versus internally generated. It is also possible that, following an initial phase of external colonisation, an internal mechanism subsequently emerged in the course of evolution, as a more efficient route for generating evolutionary innovation.

\section{Conclusions}

\if\afhead1 {\parop{recap}} \fi
ESTA frames development as a continuous process extending throughout the lifespan, driven by developmental genes that progressively reshape cellular epigenetic states. Within this framework, ageing reflects the expression of late-acting modifications that remain subject to ongoing evolutionary optimisation and function as somatic ``experiments'' exploring phenotypic novelty. In this work, we examined the potential role of transposable elements in this process. We propose that these elements contribute to the expansion and diversification of gene regulatory networks by providing transcription factor binding sites of developmental genes, while their progressive epigenetic derepression later in life may contribute to pseudo-randomness of gene expression and its ageing-associated phenotypes. This perspective integrates transposable elements into the ESTA framework and provides a conceptual basis for interpreting their dual role in evolution, development, and ageing.

\if\afhead1 {\parop{future work}} \fi
We propose that future research should prioritise the investigation of how developmental genes are implemented in biological systems and how they mediate chromatin remodelling and other epigenetic modifications in target cells. Particular attention should be devoted to understanding the molecular mechanisms by which these genes are activated, how their activity is coordinated in space and time, and how they induce specific changes in the epigenetic landscape of target cells. A critical next step is the identification and functional characterisation of the key regulatory components underlying their activity, particularly the MOS and timer elements, which in the present framework play a central role in determining when and where change events occur. Elucidating these mechanisms is essential not only for testing and refining the proposed model, but also for enabling targeted manipulation of developmental genes for unlocking potential translational applications.

\bibliographystyle{apalike} 
\bibliography{lxagetrans}
 
\end{document}